# Low-Temperature Magnetoresistance Hysteresis in Granular $Cr_{1-x}Fe_xO_2$: Slow Relaxation of Transport-Active Magnetic Configurations

E. Yu. Beliayev*,·V. A. Horielyi, I. G. Mirzoiev, V. V. Andrievskii

B. Verkin Institute for Low Temperature Physics and Engineering, National Academy of Sciences of Ukraine, Kharkiv 61103, Ukraine

* Corresponding author: beliayev@ilt.kharkov.ua

## ABSTRACT

Low-temperature magnetoresistance loops of compacted $CrO_2$ powders become strongly nonmonotonic when electrical transport is dominated by a small number of spin-dependent intergranular tunnelling paths. We reanalyse previously reported data for undoped $CrO_2$ and Fe-containing $Cr_{1-x}Fe_xO_2$, focusing on additional branch crossings beyond the conventional low-field hysteresis. In the Fe-containing specimen, the resistance decreases, passes through a minimum, rises over an extended field interval, and then decreases again at higher field. This nonmonotonic hysteretic feature is strongly reduced when the magnetic-field sweep is slowed. Digitization of the three fastest sweeps shows that the field at which the resistance starts to recover shifts approximately linearly with sweep rate, corresponding to an effective magnetotransport relaxation time of about 3 s. This scale is attributed not to microscopic spin flips but to the slow evolution of local moment and domain configurations at the intergranular contacts dominating the current. The observations are consistent with conventional negative tunnelling magnetoresistance superimposed on a slower resistance-increasing contribution associated with transient local magnetic disorder. Fe increases coercivity and interfacial magnetic heterogeneity while reducing the overall negative magnetoresistance. The results support a transport-weighted magnetic-reconfiguration mechanism for the unusual hysteresis shape.

## Highlights

- Low-temperature nonmonotonic magnetoresistance with repeated branch crossings is analysed in compacted $Cr_{1-x}Fe_xO_2$.
- The characteristic field shifts approximately linearly with sweep rate, yielding an effective relaxation time of $2.8 \pm 0.2$ s.
- The effective time scale is associated with the relaxation of transport-active magnetic configurations at critical tunnelling links.
- Fe enhances coercivity and interfacial magnetic heterogeneity while reducing the effective spin selectivity of intergranular transport.
- The measuring current probes and weights the critical tunnelling paths, while its influence on the relaxation rate remains unresolved.

## 1. Introduction

Chromium dioxide is a canonical ferromagnetic half-metal: only one spin channel contributes at the Fermi level in the ideal electronic structure [1] [2]. In compacted powders, individual $CrO_2$ particles are separated by native or deliberately formed dielectric surface layers. Charge transport therefore occurs through an extended network of magnetic tunnel junctions, and the conductance of an individual link depends on the relative orientation of the neighbouring magnetic moments [3] [4]. At sufficiently low temperature the distribution of interparticle resistances broadens, the number of effective current paths decreases, and the measured resistance becomes disproportionately sensitive to a small subset of links [5] [6].

The conventional magnetoresistance (MR) loop of a granular ferromagnet has two resistance maxima near the coercive fields $\pm H_{\mathrm{p}}$. The resistance is largest when the moments of the electrically active particles are most disordered and decreases as the applied field aligns them. This correspondence is often expressed qualitatively as $R(H)$ following a function of $M^2(H)$. It becomes unreliable, however, when bulk magnetometry averages over the full magnetic volume while tunnelling transport samples only a sparse, strongly weighted magnetic subensemble.

This discrepancy was already evident in compacted $CrO_2$ powders. Measurements reported in Ref. [7] revealed an additional crossing of the ascending and descending MR branches at fields slightly above $|H_{\mathrm{p}}|$. This feature was attributed to percolative low-temperature transport and field-induced switching among a small number of dominant current paths. Subsequent work established strong effects of particle shape, dielectric-shell thickness, measuring current, and magnetic texture on the loop shape [8] [9] [10] [11] [12]. The most striking behaviour was found in acicular powders below approximately 10 K: the initially rapid increase in negative MR reversed at a characteristic field, the resistance then recovered despite continued magnetization, and the usual negative MR reappeared only at still higher field.

Fe substitution made this behaviour more pronounced. In the Fe-containing $Cr_{1-x}Fe_xO_2$ specimen, the low-temperature loop displayed a broad history-dependent region, repeated branch crossings, and a systematic dependence on the field-sweep rate [10]. Decreasing the sweep rate reduced $H_{\mathrm{p}}$, the maximum attainable negative MR, the characteristic field $H_{\mathrm{max}}$, and the loop area; at the slowest reported sweep, the nonmonotonic extremum became a weak shoulder. The original publication associated the effect with slow dynamics of the electrically active magnetic subsystem, but neither the microscopic relaxing object nor a characteristic time scale was established.

Branch crossing is not intrinsically unphysical. It can occur in anisotropic magnetic systems when different metastable states are projected onto the measured component [13]. The present case is nevertheless distinct because the crossing occurs in transport, is accompanied by a nonmonotonic high-field resistance, and is strongly composition- and sweep-rate-dependent. A useful precedent is the isotropic positive MR observed near the percolation threshold in Co-$Al_2O_n$ nanocomposites [14]. There, a local competition between the field of a magnetic cluster, the anisotropy of a neighbouring grain, and the external field was proposed to increase local spin noncollinearity even while the total magnetization increased. That mechanism is not assumed to be identical in $CrO_2$, but it demonstrates that local magnetic disorder may evolve oppositely to the bulk magnetization.

Here we quantitatively reanalyse the low-temperature MR curves previously reported for the undoped [7] and Fe-containing specimens [10]. We characterize the additional branch crossings, digitize the sweep-rate-dependent positions of the resistance extrema, and estimate an effective magnetotransport relaxation time. The observed loop shape is interpreted as a nonequilibrium response of slowly relaxing magnetic configurations whose contributions to the measured resistance are selected and strongly weighted by a sparse intergranular tunnelling network. These transport-active configurations are associated with the small subset of contacts that dominate the current path. The data are consistent with a delayed resistance-increasing contribution arising from transient local magnetic disorder and amplified by percolative current weighting. They do not, however, distinguish whether the underlying microscopic process involves moment rotation, domain-wall motion, surface-spin rearrangement, or redistribution of current among competing tunnelling paths.

## 2. Experimental details and data analysis

The present study reanalyses magnetoresistance curves previously reported in Refs. [7] [10] [11] [12]. Quantitative values were extracted from the published graphs by digital curve tracing. Their uncertainties were estimated from the differences between corresponding extrema on the positive- and negative-field branches and from the finite width of the plotted curves.

The principal sweep-rate data were obtained for two compacted powders prepared by hydrothermal synthesis [10]. Specimen #1 was undoped $CrO_2$, whereas Specimen #2 was a $Cr_{1-x}Fe_xO_2$ solid solution containing 75 mmol Fe per mol Cr. Both powders consisted of prismatic particles with an approximate diameter-to-length ratio of 1:10 and were covered by a native degraded β-CrOOH-based surface layer. Fe was distributed among the $Cr_{1-x}Fe_xO_2$ bulk and an Fe-enriched surface region, $Cr_{2-2x}Fe_{2x}O_3$ inclusions, and states associated with the oxyhydroxide barrier [10].

Resistance was measured by a four-terminal dc method in the Ohmic regime, normally at 100 μA. The samples were pressed parallelepipeds of dimensions approximately 12 × 5 × 3 mm$^3$, with a potential-contact separation of approximately 8 mm. Magnetic-field cycles were recorded in a Kapitza electromagnet up to approximately 1.5 T. The standard sweep rate was about 0.021 T·s$^{-1}$; additional measurements were performed at several other sweep rates. The acquisition chain included a Keithley nanovoltmeter, a Keithley 2000 multimeter, and an H-301 chart recorder. Their response times were well below the several-second scale obtained below, and no deliberately introduced multi-second averaging was used. All sweep-rate curves were recorded at the same measuring current.

**Table 1** summarizes the sample properties relevant to the present analysis. Fe increased the room-temperature coercive force from 522 to 761 Oe while reducing both the sample resistance and the magnitude of the low-temperature negative MR [10].

**Table 1.** Properties of the powders used for the principal sweep-rate comparison (room-temperature magnetic data from Ref. [10]).

| Specimen | Composition | $H_c$ (Oe) | $M_{max}$ (A m² kg⁻¹) | $M_{res}$ (A m² kg⁻¹) | $d_{eff}$ (nm) | Particle shape |
|---|---|---|---|---|---|---|
| 1 | $CrO_2$ | 522 | 83.9 | 36.6 | 24 | Prismatic, ~1:10 |
| 2 | $Cr_{1-x}Fe_xO_2$; 75 mmol Fe/mol Cr | 761 | 75.3 | 34.6 | 34 | Prismatic, ~1:10 |

The applied magnetic field is reported as $\mu_0 H$ in Tesla, and the field-sweep rate is defined by $v = |d\mu_0 H/dt|$. The absolute field position of the intermediate-field resistance minimum was determined from the positive- and negative-field branches and fitted by

$$|\mu_0 H_{min}^{(R)}| = (\mu_0 H_{min}^{(R)})_0 + \tau_{MT} \times v. \quad (1)$$

A linear fit to the three highest sweep rates gives $\tau_{MT} \approx 2.8$ s, which is used below as an effective magnetotransport relaxation time.

## 3. Results

### 3.1. Temperature evolution of the low-field MR hysteresis

**Figure 1** shows the low-field magnetoresistance of a compacted $CrO_2$ powder composed of acicular particles. At 20 K, each branch exhibits a single broad resistance maximum near $\pm H_p$, followed by a monotonic decrease in resistance with increasing $|\mu_0 H|$. At 10 K, the maxima broaden and additional structure develops in the separation between the ascending and descending branches. At 4.4 K, several extrema appear within a relatively narrow field interval, the branches cross repeatedly, and an inner subloop with the opposite circulation forms within the broader tunnelling-MR hysteresis.

These changes cannot be described solely by an increase in coercivity upon cooling or by a rigid horizontal shift of an otherwise unchanged loop. The appearance and sequence of the extrema also change, indicating that the resistance depends on slowly evolving internal magnetic configurations as well as on the instantaneous applied field. This complex hysteresis develops in the low-temperature regime of activated and strongly inhomogeneous transport, where the total resistance is dominated by a limited number of critical conducting paths [7] [11].

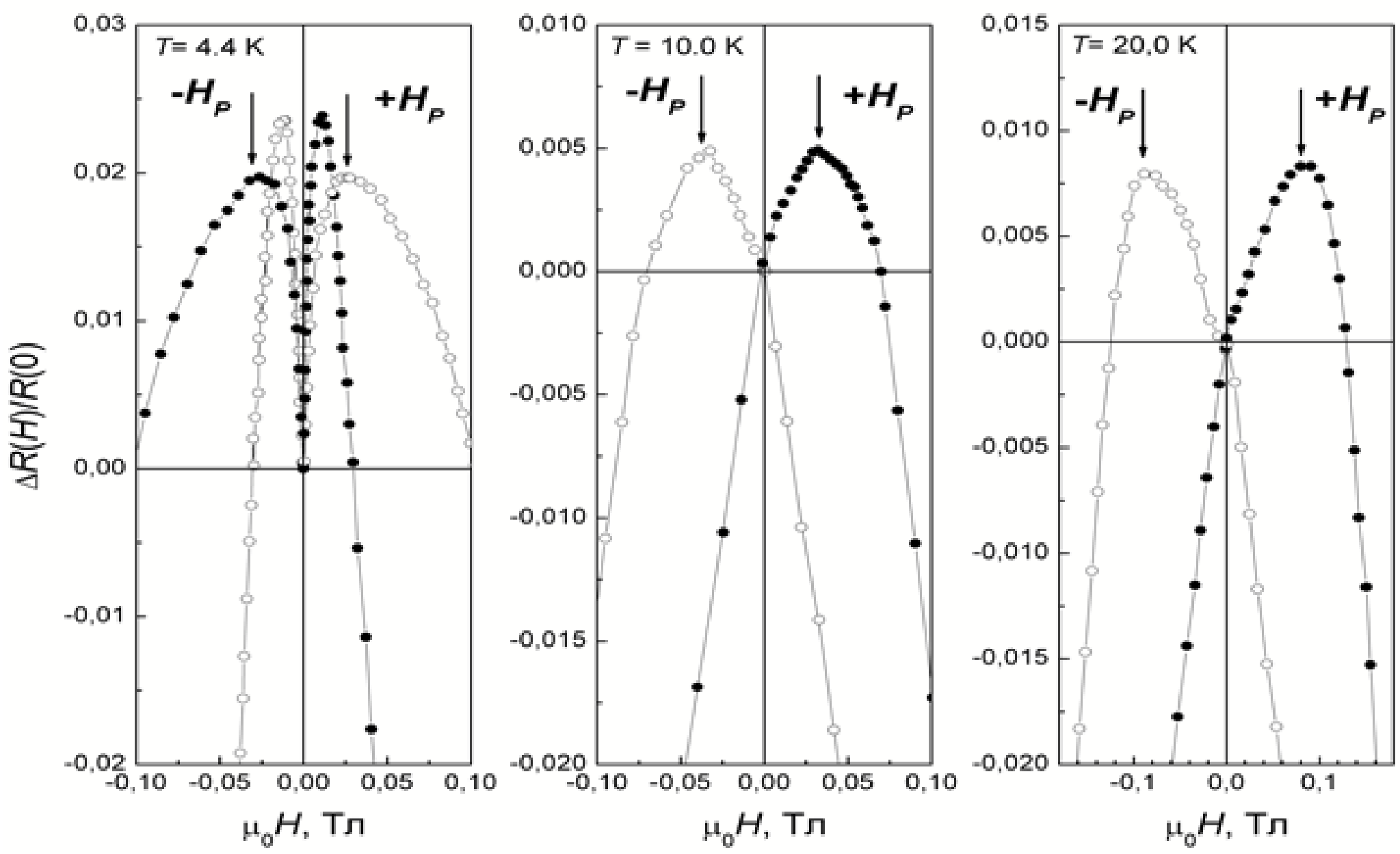

**Figure 1.** Low-field magnetoresistance hysteresis of a compacted $CrO_2$ powder composed of acicular particles at 4.4, 10.0, and 20.0 K. Filled and open symbols denote opposite field-sweep directions. At 4.4 K, several extrema and repeated branch crossings replace the conventional pair of broad resistance maxima. Data from Ref. [7].

### 3.2. Comparison of undoped and Fe-containing powders

**Figure 2** compares the undoped and Fe-containing powders at 4.25 K. Both specimens exhibit the conventional low-field resistance maxima near $\pm H_p$, together with an additional crossing of the ascending and descending branches at somewhat higher fields. In the Fe-containing specimen, however, the magnetoresistance becomes strongly nonmonotonic over a much wider field interval extending towards 1.5 T. After the resistance reaches an intermediate-field minimum, it increases again with increasing $|\mu_0 H|$. At still higher fields, the two branches approach each other and the resistance begins to decrease once more. The intermediate-field region is therefore characterized by a pronounced history-dependent resistance recovery.

The contrast between the two specimens argues against a purely instrumental origin of this behaviour. Both were measured using the same experimental arrangement, whereas the Fe-containing powder displays substantially stronger nonmonotonicity, additional branch crossings, and a much more pronounced high-field resistance recovery. Instrumental response may affect the apparent positions and widths of individual features, but it does not readily explain their strong dependence on Fe content.

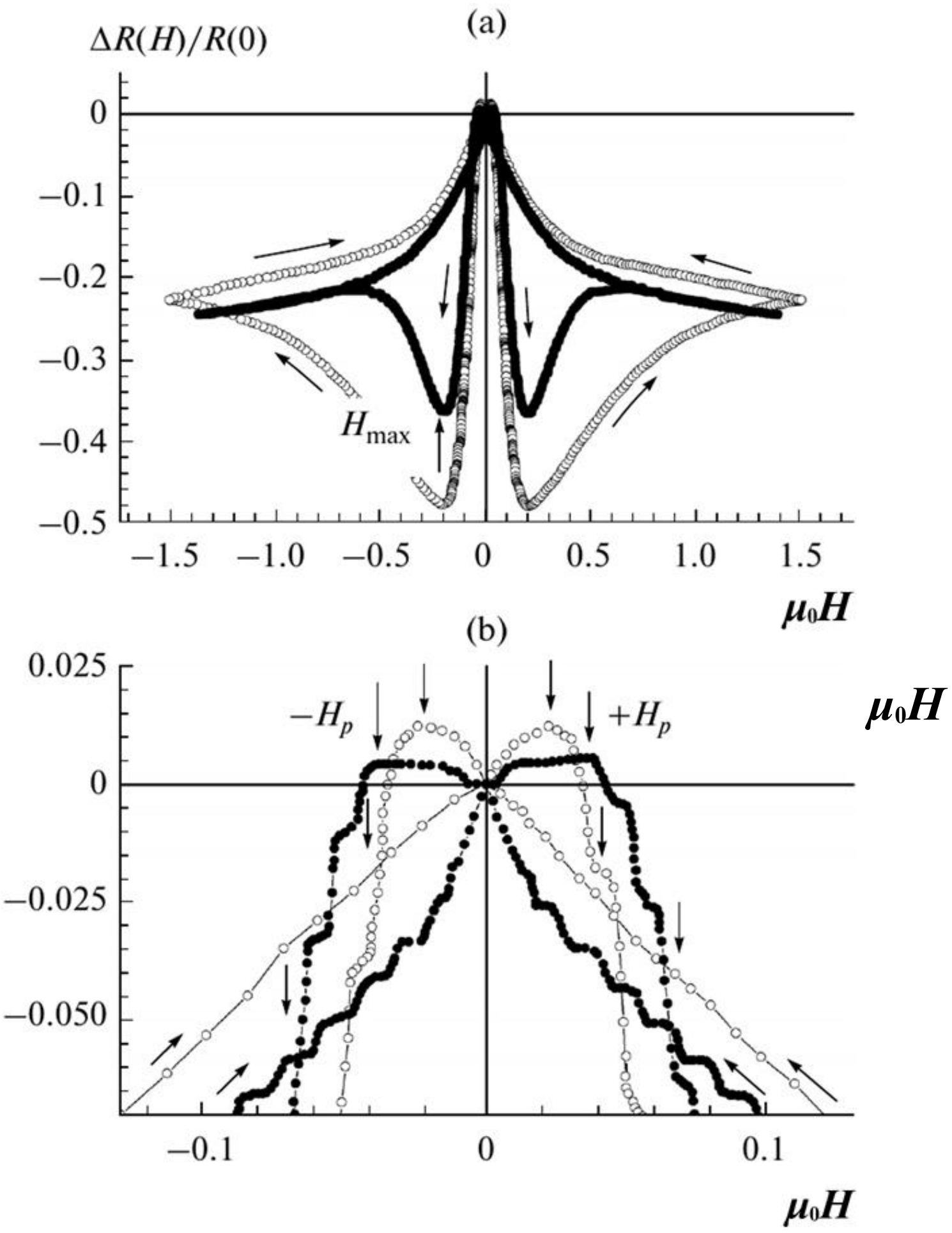

**Figure 2.** Magnetoresistance loops of undoped $CrO_2$ (open symbols) and Fe-containing $Cr_{1-x}Fe_xO_2$ (filled symbols) at 4.25 K and a field-sweep rate of approximately 0.021 $T·s^{-1}$. (a) Full field range; (b) enlarged low-field region around the resistance maxima at $\pm H_p$. Data from Ref. [10]. $\mu_0 H$

### 3.3. Sweep-rate dependence

**Figure 3** shows the magnetoresistance loops of the Fe-containing specimen measured at different magnetic-field sweep rates. At $\upsilon$ = 0.25 T $s^{-1}$, the resistance reaches a pronounced minimum only at fields close to 1 T and then increases again over a broad field interval. As the sweep rate is reduced, the minimum shifts towards lower $|\mu_0 H|$, becomes less pronounced, and the separation between the branches decreases. At the lowest rate, $\upsilon$ = 0.0029 $T·s^{-1}$, only a weak shoulder remains and the subsequent resistance recovery is barely resolved.

The simultaneous changes in the position and depth of the resistance minimum, the branch ordering, and the loop width cannot be described by a simple rigid displacement of an otherwise unchanged curve. Instead, they indicate that the measured resistance contains a slowly responding contribution whose magnitude increases when the field is swept more rapidly. The lowest-rate curve is closer to the stationary field-dependent response, whereas faster sweeping produces a pronounced nonmonotonic hysteresis.

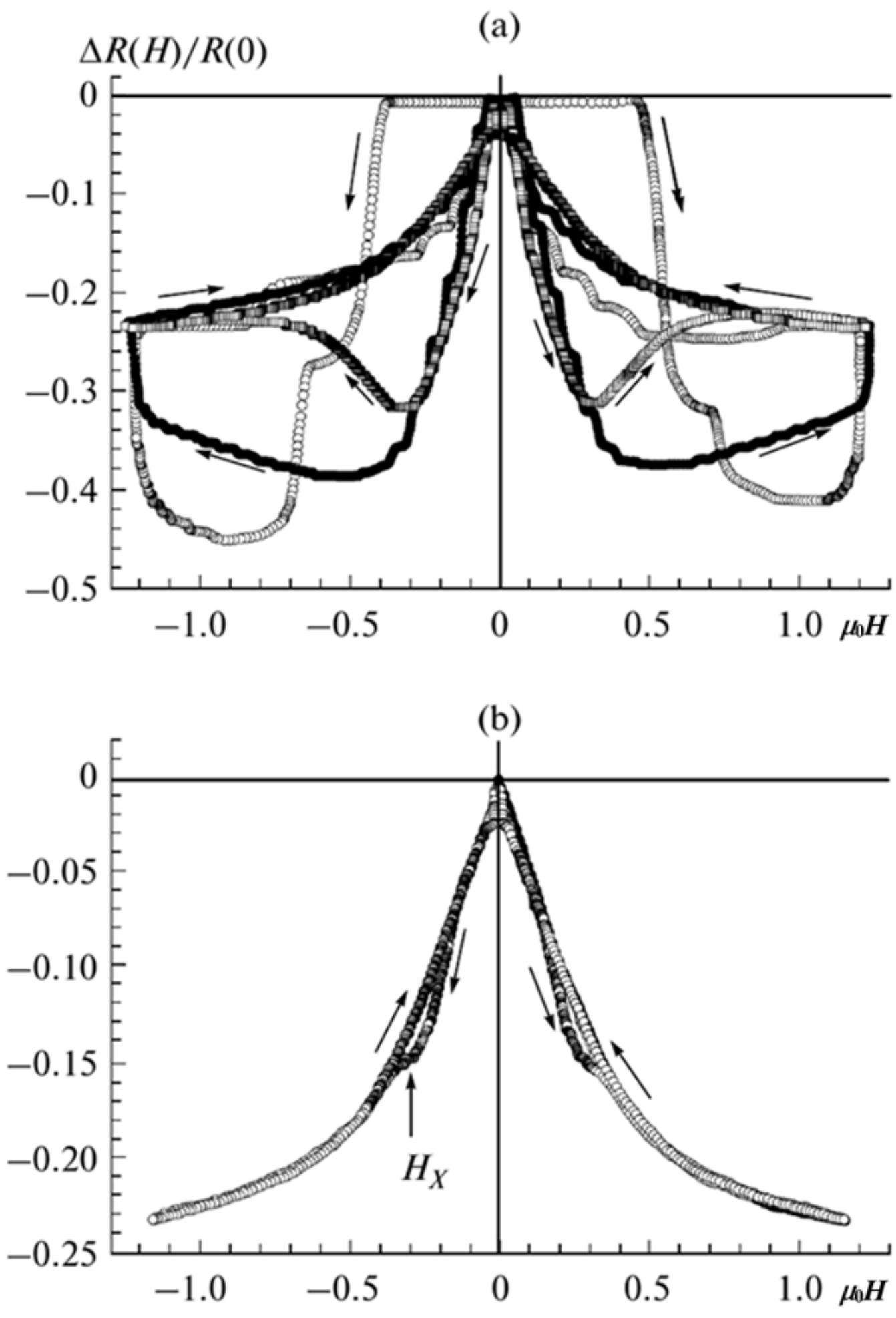

**Figure 3.** Sweep-rate dependence of the magnetoresistance hysteresis in the Fe-containing $Cr_{1-x}Fe_xO_2$ powder at 4.25 K and a measuring current of 100 μA. (a) $\upsilon = 0.25$ T·s$^{-1}$ (open circles), 0.125 T·s$^{-1}$ (filled circles), and 0.037 T·s$^{-1}$ (open squares); (b) $\upsilon = 0.0029$ T·s$^{-1}$. At the lowest sweep rate, the intermediate-field resistance minimum is reduced to a weak shoulder and the subsequent resistance recovery is barely resolved. Data from Ref. [10].

### 3.4. Sweep-rate estimate of an effective magnetotransport relaxation time

The field positions of the intermediate-field resistance minima obtained from the positive- and negative-field branches are listed in **Table 2**. Their mean absolute values shift systematically towards higher fields as the sweep rate increases. This dependence is described empirically by Eq. (1), and the resulting weighted linear fit is shown in **Figure 4**. A fit to the three data points for which the minimum is clearly resolved gives $(\mu_0 H_{min}^{(R)})_0 = (0.2 \pm 0.03)$ T, $\tau_{MT} = (2.8 \pm 0.20)$ s. Since the slope relates a field displacement to the field-sweep rate, it has the dimensions of time. We interpret $\tau_{MT}$ as an effective magnetotransport relaxation time associated with the slow evolution of transport-active magnetic configurations rather than with microscopic electron-spin relaxation. For comparison, microscopic electron phase and spin-related relaxation times extracted from weak-localization magnetoconductivity in strongly disordered Au films lie in the $10^{-12} – 10^{-10}$ s range [15]. The present several-second scale must therefore characterize a collective reconfiguration of transport-critical magnetic contacts rather than electronic decoherence or an elementary spin-flip process. Because the estimate is based on three values extracted from the published curves, the linear relation is used as an empirical description of the observed sweep-rate dependence, not as evidence for a unique microscopic relaxation law.

**Table 2.** Field positions of the intermediate-field resistance minima used to estimate the effective magnetotransport relaxation time.

| $v$ (T s$^{-1}$) | $\mu_0 H^{(R)}_{min,-}$ (T) | $\mu_0 H^{(R)}_{min,+}$ (T) | $\lvert\mu_0 H^{(R)}_{\text{min}}\rvert$ (T) | Estimated $\sigma_H$ (T) |
|---|---|---|---|---|
| 0.250 | -0.884 | +0.988 | 0.936 | 0.052 |
| 0.125 | -0.539 | +0.545 | 0.542 | 0.020 |
| 0.037 | -0.298 | +0.321 | 0.310 | 0.020 |

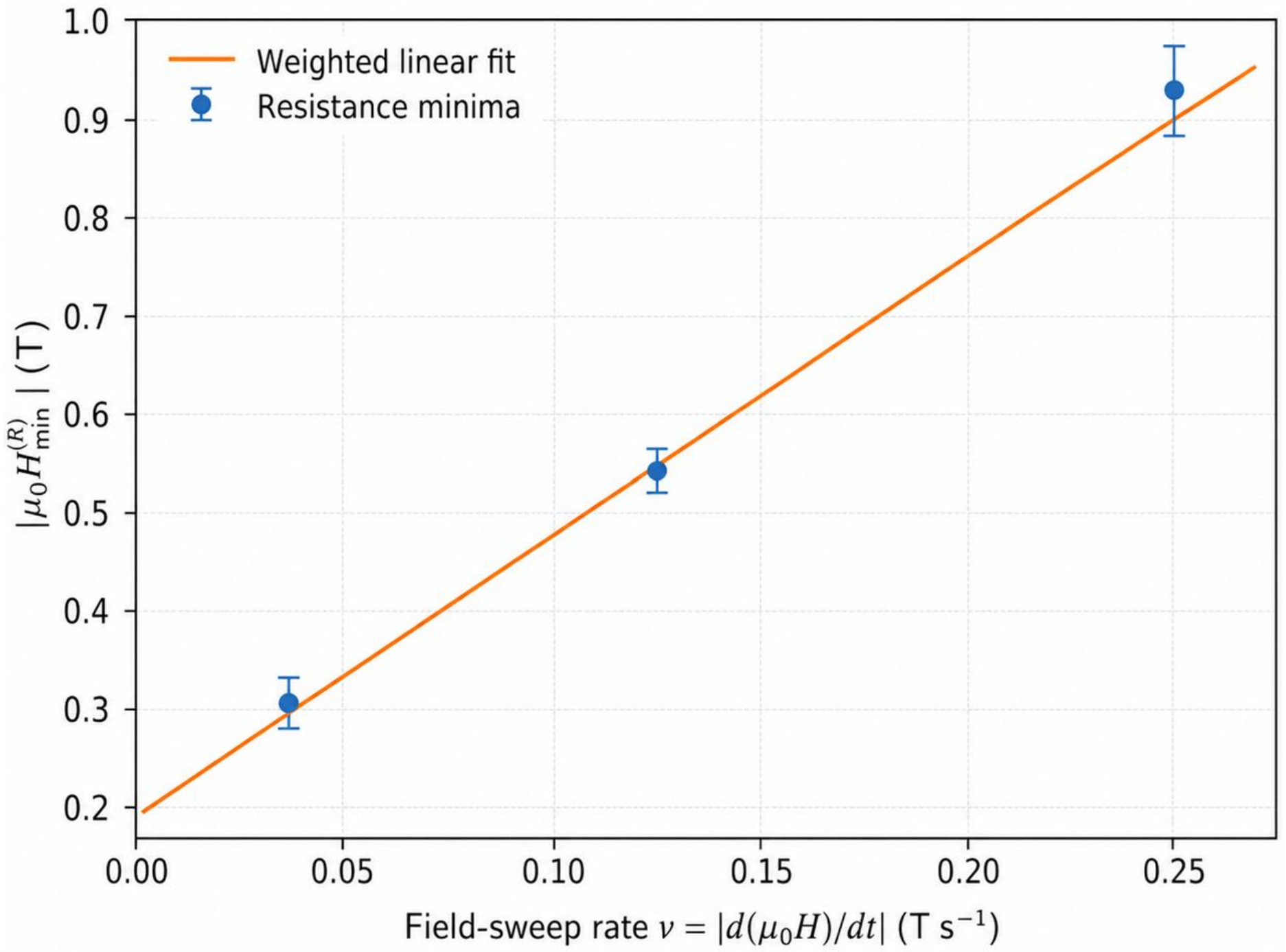

**Figure 4.** Mean absolute field position of the intermediate-field resistance minimum as a function of magnetic-field sweep rate. The error bars account for the difference between the positive- and negative-field branches and the uncertainty in reading the published curves. A weighted fit to the three data points gives $(\mu_0 H^{(R)}_{min})_0$ = (0.2 ± 0.03) T, and $\tau_{\text{MT}}$ = (2.8 ± 0.20) s. The slope is interpreted as an effective magnetotransport relaxation time.

## 4. Minimal kinetic model

### 4.1. Two transport contributions

The sweep-rate dependence cannot be represented by a simple displacement of an otherwise unchanged magnetoresistance curve, because both the loop shape and the ordering of its branches change. We therefore describe the normalized magnetoresistance as

$$m(t) = \frac{\Delta R(t)}{R(0)} = m_{bg}[\mu_0 H(t), \mathcal{B}] + m_{slow}(t). \quad (2)$$

where $m_{bg}$ represents the conventional negative tunnelling magnetoresistance associated with field-induced alignment of the magnetic moments contributing to electrical transport. The symbol $\mathcal{B}$ labels the field branch and the preceding magnetic-field history, allowing the background contribution to remain hysteretic.

The second term, $m_{slow}$, is a resistance-increasing contribution associated with the gradual reconfiguration of the small subset of intergranular contacts that dominate the current path. It may reflect changes in relative moment orientation, domain structure, or interfacial magnetic disorder, together with the resulting redistribution of current among competing tunnelling paths. The present data do not distinguish among these microscopic processes. A schematic representation of such a transport-critical contact is shown in **Figure 5(a)**.

At fixed temperature and measuring current, the field-dependent stationary value of this contribution is denoted by $m_{slow}^{st}(\mu_0 H)$. Its evolution is described phenomenologically by

$$\tau_{\mathrm{MT}} \frac{dm_{slow}}{dt} = m_{slow}^{st}\,[\mu_0 H(t)] - m_{slow}(t) \tag{3}$$

Eq. (3) is the simplest phenomenological description containing a single characteristic time. It is not intended as a microscopic relaxation law for an individual spin or domain wall. Instead $\tau_{\mathrm{MT}}$ characterizes the collective transport response of the critical contacts and is associated with the several-second scale obtained from the sweep-rate dependence in Section 3.4.

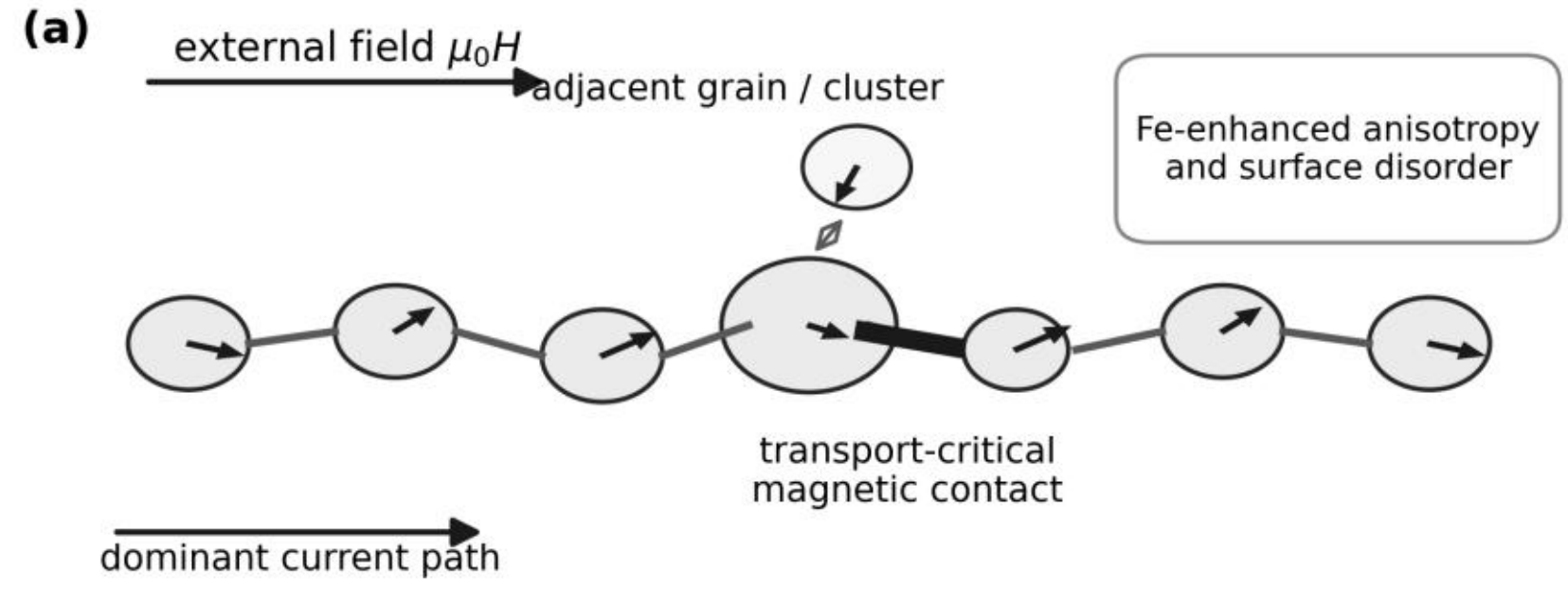

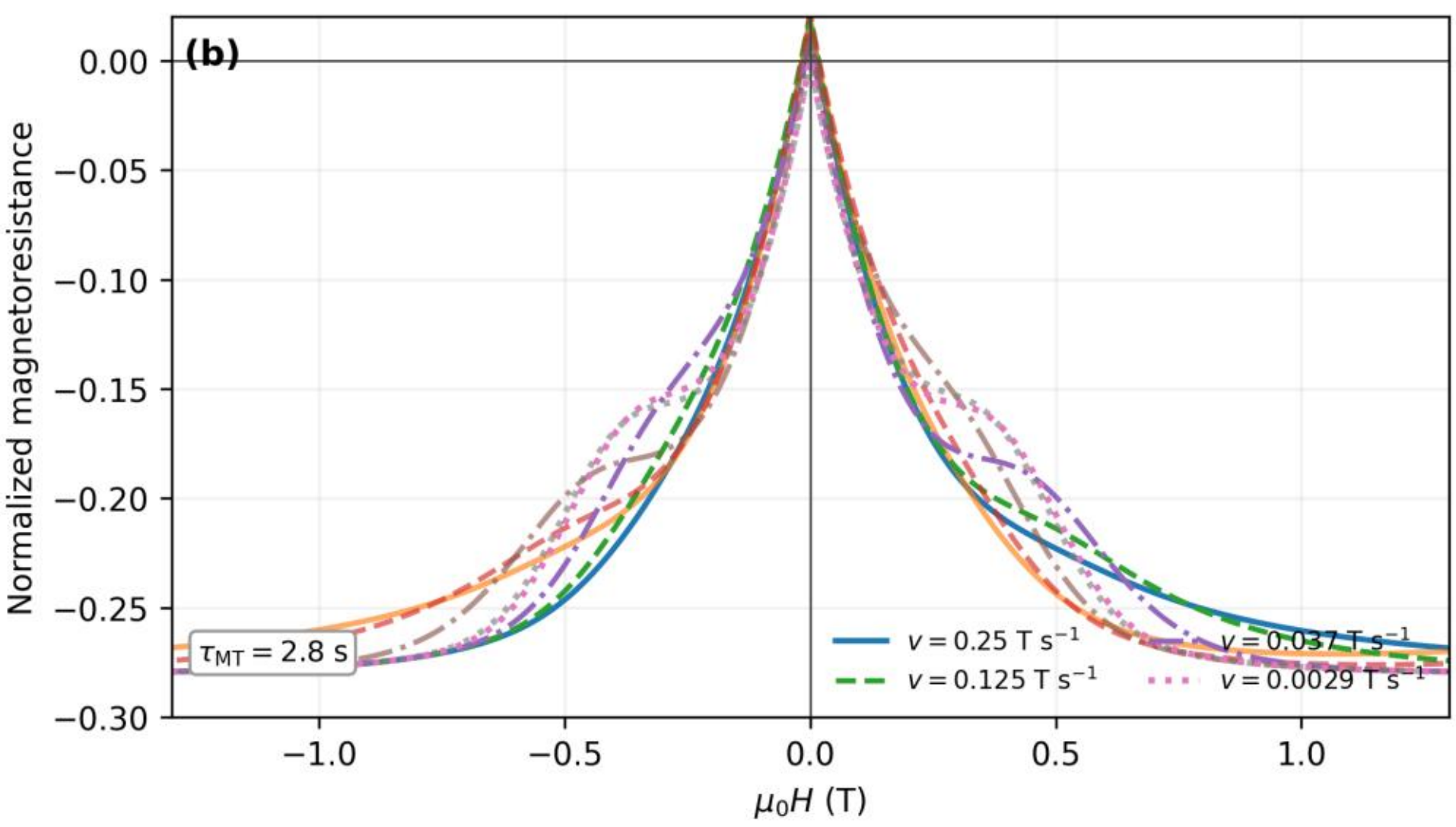

**Figure 5.** Phenomenological kinetic picture and illustrative calculation. (a) Schematic representation of a sparse current path containing a transport-critical magnetic contact whose local configuration may lag behind the applied field because of interfacial disorder and coupling to neighbouring grains or clusters. Fe-related surface disorder may broaden the distribution of local magnetic barriers. (b) Illustrative solution of Eqs. (2) and (3) for $\tau_{\mathrm{MT}} = 2.8$ s, combining a conventional negative tunnelling magnetoresistance background with a delayed positive contribution from local magnetic disorder. The calculation reproduces sweep-rate-dependent branch crossing, intermediate-field resistance recovery, and collapse toward a weak shoulder in the slow-sweep limit. It is intended as a qualitative illustration rather than a microscopic fit.

### 4.2. Possible microscopic origin of the slow contribution

A related microscopic mechanism was proposed for granular Co–$Al_2O_n$ near the percolation threshold [14]. In that system, a magnetically hard cluster can produce a local dipolar field acting on a neighbouring softer grain. An applied field may rotate the softer moment while the cluster remains blocked, temporarily increasing their relative angle and the resistance of the contact between them. At higher field, both moments become more closely aligned and the resistance decreases again. Because such a contact may represent only a very small fraction of the magnetic material, its transport signature need not be accompanied by a comparable feature in the bulk magnetization.

A similar local process may occur in compacted $CrO_2$ powders, although the present data do not identify a unique microscopic motion. Possible contributions include relative rotation of neighbouring moments, displacement of domain walls in larger particles, and rearrangement of interfacial spins. At low temperature, these processes may become slow on the time scale of the field sweep. Their effect on the measured resistance is amplified because the current is concentrated within a sparse network of intergranular tunnelling paths. Such enhanced sensitivity to a limited number of conducting links is a general feature of granular ferromagnets close to the percolation threshold [16] [17].

This transport weighting can be expressed using Cohn’s theorem for a linear resistor network: the sensitivity of the total resistance to a particular bond is proportional to the square of the current carried by that bond [18]. Consequently, magnetic reconfiguration at a contact belonging to a critical current path can produce a measurable change in the total resistance while remaining almost invisible in volume-averaged magnetic measurements. In this sense, *transport-active* refers to the strong electrical weighting of a small magnetic subensemble; it does not imply that the measuring current necessarily drives its motion. This interpretation is consistent with the transport-weighted coercivity framework recently developed for granular $CrO_2$ [19].

Fe incorporation increases the coercivity and modifies the surface and intergranular regions of the particles [10]. These changes are expected to broaden the distribution of local magnetic barriers and tunnelling conductances, thereby increasing the likelihood that a slowly responding magnetic contact also becomes important for the current path. At the same time, the smaller negative MR of the Fe-containing specimen indicates reduced effective spin selectivity of intergranular transport. The stronger dynamic feature is therefore more naturally associated with Fe-induced magnetic and interfacial heterogeneity, amplified by percolative current weighting, than with an increased spin polarization of the tunnelling current.

### 4.3. Kinetic origin of branch crossing

For a magnetic-field ramp with $d(\mu_0 H)/dt = \pm\upsilon$, the solution of Eq. (3) can be expanded to first order in $\tau_{\mathrm{MT}}\upsilon$:

$$m_{\mathrm{slow}} \simeq m_{\mathrm{slow}}^{\mathrm{st}} \mp \tau_{\mathrm{MT}}\upsilon \frac{dm_{\mathrm{slow}}^{\mathrm{st}}}{d(\mu_0 H)}. \tag{4}$$

The difference between the increasing- and decreasing-field branches is therefore

$$\Delta m_{\mathrm{branch}} \equiv m_{\uparrow} - m_{\downarrow} \simeq -2\tau_{\mathrm{MT}}\upsilon \frac{dm_{\mathrm{slow}}^{\mathrm{st}}}{d(\mu_0 H)}. \tag{5}$$

The branch difference changes sign when the derivative of the stationary slow contribution changes sign. A localized resistance-increasing contribution can therefore produce additional branch crossings and an inner subloop with circulation opposite to that of the broader tunnelling-MR hysteresis. The finite relaxation time also displaces the resistance minimum by a field interval of order $\tau_{\mathrm{MT}}v$, consistent with the sweep-rate dependence obtained in Section 3.4.

**Figure 5(b)** shows an illustrative numerical solution of Eqs. (2) and (3) using $\tau_{\mathrm{MT}} = 2.8$ s and the experimental sweep rates. At low sweep rate, the calculated response retains only a weak shoulder. Faster sweeping delays the resistance-increasing contribution, allowing a deeper minimum to develop before the resistance recovers. The calculation is intended to demonstrate that the proposed kinetic mechanism can generate the observed sequence of features; it is not a fit to the experimental curves.

### 4.4. Role of the finite measuring current

The sweep-rate series was measured at a fixed dc current of 100 µA. The present data therefore do not determine whether the effective relaxation time depends on current and do not establish that the measuring current drives the observed relaxation.

In the Ohmic regime, the term *transport-active* refers to the electrical weighting of intergranular contacts within the conducting network. Contacts carrying large local currents make disproportionately large contributions to the measured resistance, but this does not imply that the current causes their magnetic reconfiguration.

More generally, both the stationary response and the effective relaxation time may depend on field, temperature, and measuring current:

$$m_{\mathrm{slow}}^{\mathrm{st}} = m_{\mathrm{slow}}^{\mathrm{st}}(\mu_0 H, I, T), \tau_{\mathrm{MT}} = \tau_{\mathrm{MT}}(\mu_0 H, I, T). \tag{6}$$

Previous measurements on granular $CrO_2$ showed that the measuring current can modify transport-derived coercive fields [20]. Local Joule heating, spin-transfer effects, and current-induced redistribution among competing tunnelling paths may therefore perturb the transport-active magnetic configurations. Their contributions cannot, however, be separated using measurements performed at a single current.

The stronger dynamic structure in the Fe-containing specimen is accompanied by a smaller negative MR and hence weaker effective spin selectivity of intergranular transport. It is therefore unlikely to result simply from an increased spin polarization of the measuring current. Establishing a current-assisted relaxation mechanism would require sweep-rate or field-stop measurements performed at several current values.

## 5. Alternative interpretations and scope of the model

**Instrumental response.** A finite response time of the measurement system may contribute to the apparent displacement and smoothing of the magnetoresistance curves. A purely instrumental origin is nevertheless unlikely, because the same experimental arrangement produced a much weaker nonmonotonic response in undoped $CrO_2$. The pronounced composition dependence therefore indicates that the dominant contribution originates in the specimen. The extracted $\tau_{\mathrm{MT}}$ should nevertheless be regarded as an effective time scale that may include a smaller instrumental contribution.

**Magnetic relaxation.** A distribution of anisotropy barriers and the associated magnetic viscosity are likely to contribute to the sweep-rate dependence. However, the electrical resistance does not represent a volume average of the magnetic state: it is strongly weighted towards a small number of critical tunnelling contacts. The observed branch crossings and resistance recovery can therefore reflect local magnetic relaxation even when the corresponding feature is weak or absent in the bulk magnetization. Magnetic relaxation and percolative transport weighting should thus be regarded as complementary parts of the same mechanism.

**Other magnetotransport contributions.** Conventional anisotropic magnetoresistance is not expected to dominate because the principal feature is nearly symmetric in field, develops only in the low-temperature inhomogeneous-transport regime, and changes strongly with sweep rate. Hall-voltage admixture would contribute mainly an odd-in-field component, whereas the nonmonotonic structure discussed here is approximately even. Small contributions from either effect cannot be excluded, but they do not naturally account for the main observations.

**Scope and limitations.** The present data do not uniquely identify the microscopic origin of the slow contribution. Relative moment rotation, domain-wall motion, interfacial-spin rearrangement, and redistribution of current among competing tunnelling paths may all contribute. Measurements at a single current also cannot determine whether the current merely probes the transport-active configurations or weakly modifies their relaxation.

The estimate of $\tau_{\mathrm{MT}}$ is based on three sweep rates for which the intermediate-field resistance minimum is clearly resolved. It should therefore be interpreted as an effective characteristic time rather than a complete relaxation spectrum. Measurements combining sweep-rate variation, field-stop protocols, current dependence, and simultaneous magnetization and resistance would be required to distinguish the possible microscopic processes.

## 6. Experimental tests of the proposed interpretation

The proposed interpretation leads to several experimentally testable expectations. If the magnetic field is held constant near the intermediate-field resistance minimum or within the resistance-recovery region, the resistance should evolve towards the corresponding slow-sweep value on a time scale of several seconds. The direction and magnitude of this relaxation should depend on the holding field and the preceding sweep direction. At sufficiently low sweep rates, the separation between the ascending and descending branches is expected to vary approximately linearly with sweep rate. Simultaneous transport and magnetization measurements would test the role of percolative weighting: a much stronger relative time dependence in $R$than in the bulk $M$would indicate that the relaxation is concentrated at a small number of transport-critical contacts.

Measurements at several dc currents are needed to determine whether the current only probes the conducting network or also modifies its stationary magnetic state and effective relaxation time. Systematic variation of Fe concentration and particle-surface chemistry would provide an additional test, because the proposed mechanism assigns a central role to interfacial magnetic barriers and tunnelling states. A suitable protocol would combine field-sweep-rate variation with field holds at selected points of the nonmonotonic loop while recording the resistance, sample voltage, and applied field. Simultaneous magnetization measurements, where feasible, would help separate bulk magnetic relaxation from its transport-weighted manifestation.

## 7. Conclusions

The low-temperature magnetoresistance hysteresis of compacted $CrO_2$ powders develops repeated branch crossings and a pronounced intermediate-field resistance recovery when electrical transport is dominated by a sparse network of intergranular tunnelling paths. These features are much stronger in the Fe-containing specimen and are progressively reduced as the magnetic-field sweep rate decreases.

The field position of the intermediate-field resistance minimum shifts approximately linearly with sweep rate. Analysis of the three curves for which the minimum is clearly resolved gives an effective magnetotransport relaxation time of approximately 2.8 s. This value characterizes the slow evolution of magnetic configurations at transport-critical contacts and should not be interpreted as a microscopic electron-spin relaxation time.

The observations are consistent with a conventional negative tunnelling-magnetoresistance contribution superposed on a slower resistance-increasing response associated with local magnetic disorder. Possible microscopic processes include relative moment rotation, domain-wall motion, and rearrangement of interfacial spins, whose influence on the total resistance is amplified by percolative current weighting. Fe increases coercivity and modifies the magnetic and electronic structure of the intergranular regions while reducing the effective spin selectivity of the tunnelling current. The enhanced dynamic response of the Fe-containing powder is therefore more naturally associated with increased magnetic and interfacial heterogeneity than with stronger current spin polarization. Measurements at several sweep rates and currents, combined with field-stop and simultaneous magnetization experiments, would be required to distinguish the possible microscopic mechanisms.